# S$^4$oC: A Self-optimizing, Self-adapting Secure System-on-Chip Design Framework to Tackle Unknown Threats – A Network Theoretic, Learning Approach

Shahin Nazarian and Paul Bogdan
shahin.nazarian@usc.edu, pbogdan@usc.edu
Ming Hsieh Department of Electrical and Computer Engineering, University of Southern California, Los Angeles, CA, USA

*Abstract*— We propose a framework for the design and optimization of a secure self-optimizing, self-adapting system-on-chip (S$^4$oC) architecture. The goal is to minimize the impact of attacks such as hardware Trojan and side-channel, by making real-time adjustments. S$^4$oC learns to reconfigure itself, subject to various security measures and attacks, some of which possibly unknown at design time. Furthermore, the data types and patterns of the target applications, environmental conditions, and sources of variations are incorporated. S$^4$oC is a manycore system, modeled as a four-layer graph, representing the model of computation (MoCp), model of connection (MoCn), model of memory (MoM) and model of storage (MoS), with a large number of elements including heterogeneous reconfigurable processing elements in MoCp, and memory elements in the MoM layer. Security driven community detection, and neural networks are utilized for application task clustering, and distributed reinforcement learning (RL) for task mapping.

## I. Introduction

To cope with increasing complexity of data and applications as well as energy consumption related issues, the trend has been to utilize multi/manycore platforms for modern architectures. Such architectures are traditionally designed using a combination of full-custom, semi-custom ASIC (Application Specific Integrated Circuits) and FPGA components with area, performance or energy as the main goal. They mostly include homogeneous components with hardware and software solutions separately designed and added to them. E.g., to exploit the high degree of parallelism available on such platforms, programmers typically write multithreaded applications where multiple threads are spawned and mapped onto the cores by the operating system (OS). These threads share data by storing one copy of the data into the data cache. If one thread updates the shared variable, the same new copy must be broadcast to the rest of cores for consistency. Network-on-Chip (NoC) has been proven effective for data transfer among cores. For each core, i.e., a processing element (PE), there exists a router, i.e., a connecting element (CE) to transfer data from one node to another and the CE keeps track of the next node to which data should be sent. However, there are drawbacks associated with such platforms: (1) Each core may be designed to be simple, general, and possibly homogeneous. In addition to the performance and power dissipation issues related to such cores, the simplicity, and homogeneity of such platforms and their data movements, creates vulnerabilities that could jeopardize the security of functionality and information [1]-[10], and (2) Countermeasures to security attacks may not be sufficient in case those threats evolve.

We envision a systematic design, verification and optimization flow for manycore systems from top system level behavioral view, down to layout (i.e. spec to gdsii), which is driven by security goals and performance and energy constraints. This flow would deliver a system-on-chip (SoC) that learns to reprogram and reconfigure itself even in real-time in the event of new security threats. Our <u>s</u>elf-optimizing, <u>s</u>elf-adapting <u>s</u>ecure <u>SoC</u>, called S$^4$oC, provides flexibility, programmability, and performance/energy efficiency that are demanded for complex applications in heterogeneous systems [11]-[20]. S$^4$oC is domain specific and uses various static and real-time approaches to address security. We construct a four-layer security-aware graph model of the S$^4$oC system, namely computation, connection, memory and storage layers to capture the task activities while they are mapped onto the S$^4$oC elements. At compile time, the high-level programs of the target applications are converted to our low-level virtual machine (LLVM) intermediate representation (IR) instruction set, which is LLVM IR compatible, with plugins accounting for security. Next a hierarchical task graph model is constructed from the LLVM IR instructions accounting for task dependencies. An offline learning process using NN classifiers is applied to learn special features related to functionality and security including unknown effects. Community detection (CD) is used to partition the hierarchical task graph into non-overlapping clusters. Considering the dynamic, uncertain nature of data, hardware, and security threats, the runtime steps of task scheduling and resource reconfiguration are performed by distributed RL based scheduler agents. In the following, we present a summary of common security threats and their countermeasures. We then present the components and flow of S$^4$oC (cf. Figure 1).

## II. Background

This section presents a summary of common hardware security threats and countermeasures.

### A. Hardware Attacks

*Side-Channel*: The adversary monitors certain design parameters such as those related to power, thermal, latency, and electromagnetic radiation behavior of the system, from which it attempts to gain some secret information, e.g., about the implementation of the system (such as the CNN model [9]) or certain private keys. It closely observes the primary and side-channel outputs and examines the variations in the side-channel and primary inputs. It aims to correlate those inputs and outputs to be able to derive the secret information [1]-[10]. Due to implicit and at times unknown nature of side-channel attacks, they are extremely efficient in evading present-day hardware security techniques.

*Hardware Trojan*: Malicious modifications to an existing circuit with the goal of changing or controlling the functionality of the circuit, stealing its secret information, degrading its reliability, performance, etc., or result in denial-of-service [20]-[32]. They may be added to the circuits during any phase of the design cycle, including fabrication, or earlier, by manipulating the RTL code or modifying the macros during RTL design and synthesis. Also, EDA tools may contain malicious code that collects secretive data in IPs (Intellectual Properties).

*IP Piracy*: Any illegal form of using an IP, IC or SoC. Including stealing and claiming ownership [33]-[36]. It typically manifests itself in the form of overbuilding and illegally selling an IP. The malicious foundry can overbuild ICs without the consent of the designer and sell the excess ICs in the gray market. Counterfeiting is another form of attack where the attackers illegally imitate the original design.

*Reverse Engineering*: A process which aims at identifying and understanding the design, functionality and structure of the integrated circuit [37]-[42]. Reverse engineering of integrated circuits has been a common practice in design verification and commercial piracy investigation. However, the technique can be misused by the adversary for the purpose of stealing and even pirating the design. For instance, starting from the design layout, the adversary can obtain the transistor level netlist and ultimately the gate level netlist. They can therefore use the netlist to build their own circuit or sell the stolen design illegally. In case of an unknown module, the functionality can be identified based on Boolean satisfiability checks or behavioral comparison to known components.

### B. Countermeasures

The following is a brief overview of the solutions that may address one or more of the security threats.

*Physical Unclonable Functions*: Cost-efficient cryptographic primitives that exploit the sources of variations in the chip components (e.g., the SRAM modules). PUFs generate random values that would be unpredictable even for an attacker with physical access to the system. Among the sources of variation, environmental factors such as temperature, supply voltage and electromagnetic interference, as well as circuit aging impacts the quality of PUFs. It is therefore crucial for a PUF to be different among different circuits while remaining robustly unchanged on the same circuit [43]-[49]. In [48] a PUF for reconfigurable circuits is presented that can be embedded into the design's HDL.

*Metering*: A set of security mechanisms and protocols to provide designers with post fabrication capability of tracking their design, by monitoring the number of produced ICs, their properties and possibly even controlling them remotely [50]-[59]. Some of the metering techniques are based on PUFs. Metering can be classified into active and passive techniques. Passive methods could be looked at as indenting of serial numbers on each device physically. Active metering equips the designer with not only product identification, but also control.

*Multi-Valued Numeral Systems*: To make it computationally expensive for the adversary to identify private keys by computing in non-binary or mixed-radix. Prior work has shown how a mixed binary and ternary computing can be used to enhance security, e.g., in enhancing the security of cryptography calculations [60]-[68]. Mixed radix is also capable of enhancing the performance. A core may perform computation based on a hybrid binary, ternary, quaternary and multi-radix number systems, to explore the security vs. energy/performance tradeoffs.

*Logic Encryption and Obfuscation*: Based on adding extra circuitry to hide the functionality of the design. Adversary would need a valid key to be able to unlock the circuit for normal operation. Therefore, the goal of logic locking is to make it impossible for the attacker to identify the keys. The extra logic would then help in encrypting the original design. The key inputs are stored in an on-chip tamper-proof memory. The additional logic which is added to the original circuit may consist of logic gates, multiplexers, lookup tables, or their combinations [69]-[82]. Designers provide the secret keys post fabrication, which means attackers cannot easily derive information about the circuit through brute force attacks because in the absence of the keys, the design produces wrong results. The intentional obscuring of the design used for securing sequential circuits is typically referred to as the logic obfuscation [83]-[88]. State obfuscation can protect ICs against hardware trojans, reverse engineering and piracy.

*Watermarking*: The process of embedding certain information such as a signature into the design to be able to later trace its ownership and deter infringements [89]-[95]. Differently from hardware metering that aims for a unique tag for every chip, watermarking seeks to assign the same tag to all the chips produced from a specific design. Watermarks can be embedded in the sequential part of the IP by adding additional stages and transitions in the FSM, or by adding some special constraints at the various synthesis stages of the ASIC and FPGA design flows. Watermarking should not affect the actual functionality of the IP. Furthermore, copying or removing it should not be straightforward. Although IP watermarking serves well for proving authorship of an IP, it cannot prevent the adversary from gaining information about the IP.

*Camouflaging*: Mainly a countermeasure to adversarial reverse engineering, and typically based on designing circuits with logic cells whose layouts look alike but differ in their functionalities. Once the adversaries wrongly interpret the functionality, they will obtain netlists which are different from the original one. For example, camouflaged NAND and NOR gates are built to look the same in layout, with the type of contact, i.e., real vs dummy as the only difference. While a real contact electrically connects two layers, but a dummy contact creates a gap in between the layers and thus electrically they are not connected. Prior work [37] showed that camouflaging, using an efficient algorithm to choose a handful of gates may slow down adversarial reverse engineering. However, a combination of SEM images and applying input patterns to the circuit to generate the output, may significantly shorten the time for the adversary to identify the camouflaged gates [96].

*Split Manufacturing*: Originally developed to enhance the fabrication yield, but more recently for security purposes as well. It is simply to divide the integrated circuits netlist into multiple parts and fabricate each of them at a different foundry, so that no single foundry gets access to the complete IC design. This process typically provides higher levels of security in comparison to other countermeasures against the various hardware-based attacks [97]-[99]. Split manufacturing processes typically divide the layout into two parts, namely Front End of Line (FEOL) and Back End of Line (BEOL) each of which to be fabricated in different foundries. The FOEL layers consist of transistors and some lower metal layers and are manufactured by an untrusted foundry and the BEOL layers consist of higher metal layers and are manufactured by a trusted foundry. Therefore, an adversary in the untrusted foundry has access to a partial netlist only.

*Hardware Trojan Counteracts:* the methods are categorized as: (1) Detections [100]-[102], (2) Design for security [103]-[105], and (3) Runtime monitoring [106]-[107]. Detection is the technique that determines whether a trojan exists in a piece of IC. Design for security increases the difficulty of trojan insertion or facilitates the detection of trojans. Runtime monitor is on-chip monitoring for hardware trojans during runtime. Detecting hardware trojans is challenging, because the modularity, hierarchy and regularity of custom and programmable IC design techniques necessitate numerous soft, firm and hard IPs, all of which exposed. The shrinking of layout geometries makes trojans harder to detect. Trojans get activated by certain rare conditions, which exacerbates the detection challenge. The detection techniques need frequent updates, as they may not work for new hardware trojans.

### III. S$^4$oC Framework – Graph Modeling

We propose a hierarchical multi-layer graph model for our S$^4$oC framework. There are two sets of graphs, namely the S$^4$oC architecture graph set and the task graph set.

#### A. Multi-layer S$^4$oC Architecture Graph Set

The computation, connection, memory, storage activities and their interconnected dependencies related to the S$^4$oC architecture are constructed in four graph layers, namely the model of computation (*MoCp*), model of connection (*MoCn*), model of memory (*MoM*), and model of storage (*MoS*) layers. Each node in the four-layer graph of the S$^4$oC architecture represents a processing element (*PE*), connecting element (*CE*), memory element (*ME*) or storage element (*SE*). Each node has a *type*, e.g., the type of a PE could be CPU, GPU, PUF, ASIC or specialized HWA (hardware accelerator) for Fast Fourier Transform (FFT), matrix multiplication (MM), cryptography, homomorphic encryption, hyperdimensional computing, etc. Each node also has a parameter *available*, to record the availability level of the node (e.g., *available* = 0 means completely busy). Also, a link between two elements, e.g. a connection between a PE and a CE, is modeled by an edge. Each edge has a maximum *bandwidth* utilized as part of data transfer optimization.

#### B. Hierarchical Task Graph Set

The target applications running on S$^4$oC are modeled based on complex hierarchical graphs. At the topmost level, a graph is defined whose nodes represent the applications and edges define any inter-application dependency. The application dependency may be the result of a security measure, where two applications must cooperate with each other to enhance security. Each application node may have a preferred *logic type* which could be non-binary, to further boost security. For instance, an application may be set to run on the S$^4$oC elements that operate in ternary, quaternary or binary-coded ternary (BCD).

At a lower level, the tasks associated with each application are modeled based on an instruction dependency graph (IDG). A set of tasks is modeled as an IDG, *TS*. Nodes of *TS* represent the tasks and its edges represent messages to be transferred among the tasks. A task is modeled as a directed acyclic graph (DAG) *T*. Nodes of *T* represent the LLVM IR instructions. Data dependency between any two instructions in *T* is modeled by an edge, whose weight represents the amount of data to be transferred. Note that the IDG is the collection of the DAGs of all the applications. Each task may have a preferred execution affinity, so that it would be designated to certain PE type(s) for security and/or performance improvement purposes. Therefore, in order to improve the execution quality, tasks with a preferred execution affinity should be mapped to their most beneficial elements. For example, a task with loops may be mapped to a GPU type PE to improve performance whereas a task involving large matrices should not be sent to a CPU type PE. The two graph sets, namely the architecture graph and task graph sets can be used for analysis and optimization purposes. The S$^4$oC framework solves the following problem: *Given the architecture graph and task graph sets, cluster the tasks, map them onto S$^4$oC, and reconfigure the S$^4$oC elements, such that security is maximized, subject to certain performance and energy constraints.*

### IV. S$^4$oC Framework – Flow and Components

In this section, we describe the security driven flow and operations of S$^4$oC, which can be grouped into two stages, namely compile time resource allocation and runtime resource management and reconfiguration. Resource allocation includes application transformation, learning, and community detection (cf. Figure 1). At compile-time, S$^4$oC transforms applications into a complex graph and determines the type and location of features, logic type, etc., using neural networks (NNs). During training, sample applications are driven into the NNs to learn specific features with emphasis on the security measures, such as side-channel power, thermal, or memory activity profiles to protect the S$^4$oC IPs from adversary [9],[40],[36], perform watermarking, metering and obfuscation [51],[94] and detect any hardware trojans [20],[107]. The trained NNs have the capability of mining hidden structures in the new unseen applications. Next in the flow, is to partition the task graph into interdependent clusters with minimized data communication. In addition to energy and performance benefits, data communication minimization may further enhance security. Community detection (CD) on multi-layer graph is the main engine for partitioning. S$^4$oC manages the resources and reconfigures the elements in real-time, using security driven RL based on distributed intelligent schedulers. To optimize for security, the schedulers explore the best elements and configuration onto which heterogeneous tasks should be mapped, according to the corresponding application demands as well as the availability of computing, connecting, memory and storage elements. The heterogeneous elements of S$^4$oC include reconfigurable cores, ASIC cores, modules for HW acceleration of machine learning applications, and modules for security purposes such as cryptographic, non-binary numerical, PUF (physical unclonable function), metering, watermarking, etc., some of which may adapt in real-time, considering the existing environment and with the help of the intelligent schedulers.

S$^4$oC flow includes a verification framework that benefits from simulation, semi-formal and formal methodologies [108]-[111]. RL is used to generate all possible sequences of input test vectors needed to approach a target state as well as the corresponding path to the target state which contains a potential design bug. The design bugs could be related to circuit functionality or highlight a security loophole. This process is equipped with SAT solvers to help find the transition vectors. The verification framework also utilizes NNs to accelerate the verification of complex data paths in the target states. Certain security measures (such as side-channel related) are analyzed based on assertions and properties as part of the process. The quality of verification is measured in terms of verification coverage and time.

*A. Compile Time Security Driven Resource Allocation*

The objective is to find the optimal application task partitions considering potential attacks (side-channel, Trojan, etc.), as well as the functionalities such as loops, FFT, and MM to exploit the underlying reconfigurable heterogeneous NoC-based hardware platform. In our prior work [17] we have demonstrated the effectiveness of learning using NNs in assisting modularity-based CD to form clusters of tasks, with certain features, such as common functionality types (e.g., loops, FFT, or MM), common logic type and processing affinity. Such features are used in S$^4$oC aiming for security.

S$^4$oC first applies NNs to find and detect the special features that exist deeply in applications. Those features help identify certain tasks with specific security or functionality needs and map them to S$^4$oC elements accordingly. Our NNs attempt to learn features related to unknown security threats [112]-[129].

*A.1) Low-Level Virtual Machine (LLVM)*

LLVM is a set of compiler technologies that provides intermediate representation (IR) as a common model for program analysis, transformation, and synthesis [130],[131]. LLVM IR utilizes an architecture based on load/store instructions to transfer data, e.g., among registers, caches, and memory banks. LLVM IR based compilation provides language independence for the representation of high-level languages such as C/C++ and for building retargetable compilers. It also helps avoid low-level actions such as register spilling and function prolog/epilog insertion. It therefore provides a suitable representation for program analysis and optimization in terms of various metrics such as energy and security [132]-[134].

We model target applications by constructing the corresponding IDG task graph set. First, we execute the applications to collect dynamic LLVM IR like traces. S$^4$oC version of LLVM IR includes plugins to account for security features as part of the LLVM instructions. For instance, a hybrid of various numeral systems, such as binary, ternary and quaternary as well as multi-radix instructions are used. Second, we analyze the traces to figure out whether source registers of the current instruction depend on destination registers of the previous instructions. If data dependencies exist, we insert edges in between. Third, these instructions are profiled to get the precise data size as edge weights in IDG. Therefore, our approach combines static and dynamic program analysis. Static analysis means that we perform program dependency analysis to construct the graph. Dynamic analysis means that we run applications with representative inputs to get the dynamic IR traces because memory dependencies cannot be resolved statically. An example dynamic trace is shown below [17]:

$$\%4 = \text{and } \%2, \%3;$$
$$\%5 = \text{mul } \%2, \%4;$$
$$\%6 = \text{add } \%4, \%5;$$

The second instruction is dependent on the first instruction, as the second instruction's register %4 depends on the destination register %4 of the previous instruction. Similarly, the source registers %4 and %5 of the third instruction depend on the destination registers of the previous instructions. As shown in our SOSPCS framework [17], the time complexity of application transformation is $O(N\log D)$ where N is the number of IR instructions and D is the number of different destination registers, with D much smaller than N.

*A.2) Security Aware Learning*

During testing and simulation, we use new applications, but from the same domain as those in the training testbenches, e.g., those with cryptographic needs [135]. When new applications arrive, we first transform them into IDGs. Based on the graphs and their metrics, we prepare input data for the NNs. As part of the learning process we account for the unknown effects based on log-normal distribution, non-Gaussian statistical, and multifractal behavioral models [112]-[129]. To tame unknown threats using learning, we model a time-variant model of unknown complex malicious activities from partial observations, foreign interventions, and threat detections, e.g., to avoid revealing our learning models due to memory accesses or enhance trojan detection.

*A.3) Task Clustering*

Having detected the special features related to security and functionality, we use CD to partition the remaining graph into non-overlapping communities of tasks in order to: (1) Minimize security vulnerabilities, (2) Parallelize optimally the available S$^4$oC elements, and (3) Minimize data communication overhead. We use an iterative approach based on complex network theories that pushes the gain on a quality function Q based on the 3 goals above, while adding neighboring tasks to clusters and stop when there is no further gain. The quality gain Q is similar to the modularity Q of [17] but including security expressions to monitor design quality in terms of side-channel minimization, encryption robustness, etc.

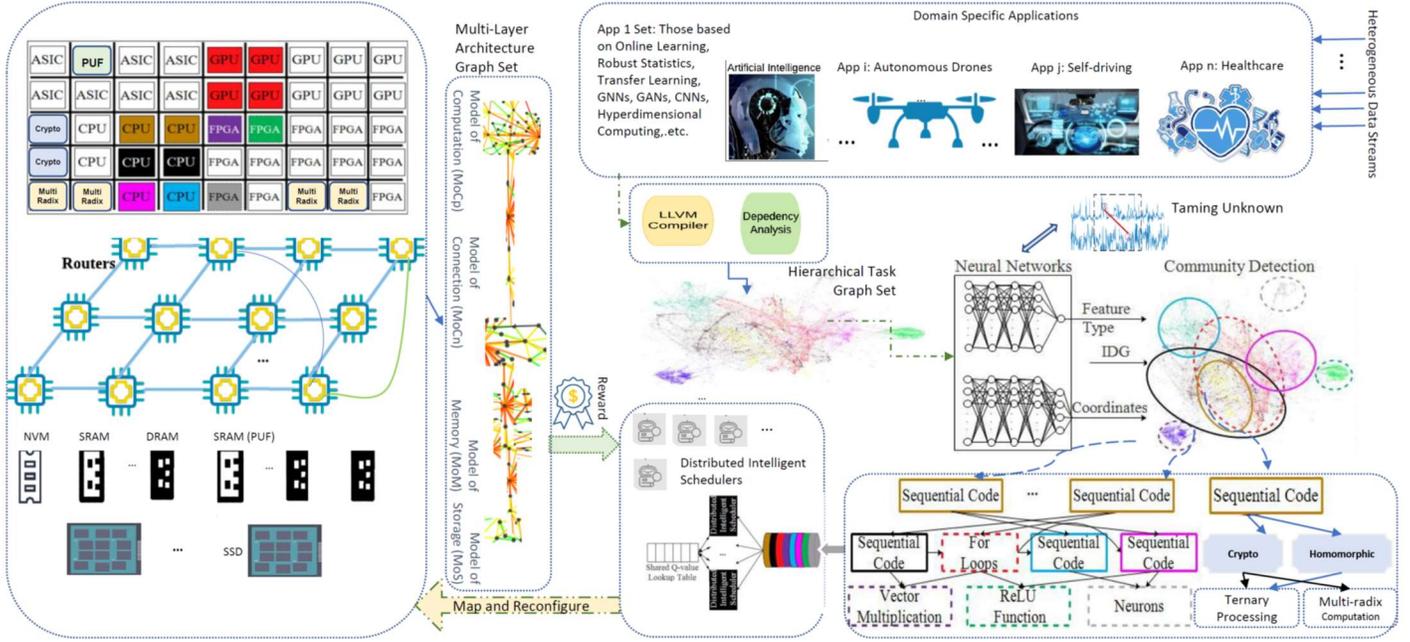

*Figure 1*: S$^4$oC Framework – Flow and Components.

### B. Runtime Scheduling and Reconfiguration

Given the task clusters, the goal is to find a mapping of those clusters onto the S$^4$oC elements such that security is maximized subject to performance and energy constraints. To gain the maximum security, the following criteria have to be ideally satisfied: (i) The preferred logic type should match that of the element(s) corresponding to the mapped tasks, (ii) Given the bandwidth of the connections, data communications should be minimized such that the mapped tasks use elements that are sufficiently close to each other.

We consider the real-time, dynamic and uncertain nature of security threats, and underlying hardware platforms such as traffic and availability of elements. Also computing, connecting, memory and storage elements must work with the heterogeneous data streams with various types and levels of security needs. Hence, traditional planning and scheduling algorithms and heuristics would not be effective in handling a runtime security driven resource planning and scheduling. We therefore design distributed RL based intelligent schedulers to map tasks to the best suitable mapping and configuration of elements of S$^4$oC. There are two advantages: (1) The technique learns and self-optimizes the mapping of tasks onto heterogeneous elements in order to gain the optimal security while meeting performance and energy constraints, (2) It is dynamic, therefore at runtime, based on the states of the environment and immediate rewards, agents decide the best policies for task mapping. In addition to the security driven design flow, S$^4$oC continues to optimize itself while at work, running applications. It employs the distributed intelligent schedulers to schedule the set of domain specific tasks and reconfigure certain hardware components.

#### B.1) Task Scheduling

Task scheduling of our framework can be modeled and conveyed into learning agents as an enhanced Balls into Bins problem [136]. In our version of the problem, the balls (tasks) need to be constantly removed from the bins (input task queues) and mapped on the S$^4$oC. Ideally agents should have a balanced workload, therefore an upper-bound is set for the workload, based on the rate of scheduled tasks.

#### B.2) Distributed RL

We believe RL with its trial-and-error nature to learn the optimal policy is very effective in runtime resource management and reconfiguration with dynamic applications [137],[138]. In our framework, states in an MDP (Markov Decision Process) environment represent the availability and type of S$^4$oC elements, and actions represent resource reconfiguration and the task mapping onto elements. A distributed Q-learning approach is used to find the optimal policy. Multiple agents interact with the environment independently. The policy is defined based on the probability of choosing a mapping given the current availability and type of elements and the goal is to optimize the quality function Q. Busy elements would be avoided by assigning large negative rewards to prevent an action to assign a task to those busy elements. During exploration, such assignments should however not be forbidden. If the target element to which a task is mapped to is busy, we perform a local search to migrate the task to the nearby element with a cost, e.g., in terms of certain constraints related to communication security, network congestion and latency.

Considering the unknown nature of some security threats and also more generally the large number of states and transitions, the system would not know all the transitions and their expected rewards. To avoid over-exploitation and getting trapped in local optimum, agents keep a balanced exploitation and exploration. For example, an ε-greedy action selection algorithm shows promising results [17]. The agent with a small probability ε randomly picks an element of S$^4$oC to which the task is to explore the possible mappings. With a high probability 1-ε, the agent chooses an element based on the Q value. All agents have access to the shared Q values and update them after receiving rewards from the environment. Hence, it is necessary to have incremental updates and approximate the quality function Q. This can be achieved by enforcing a critical section which allows only one agent access at a time.

## V. CONCLUSION

We presented S$^4$oC, a self-optimizing, self-adapting secure SoC that learns to make offline and real-time adjustments to cope with security threats, some of which might be unknown at design time. Hierarchical multi-layer graph modeling is used to model the architecture of S$^4$oC as well as tasks dependencies in the applications. More precisely, a four-layer graph, namely the model of computation, model of connection, model of memory, and model of storage layers capture the large number of elements including heterogeneous reconfigurable processing, connecting and memory elements and their interactions. Also. the target applications running on S$^4$oC are modeled based on complex hierarchical graphs to capture application and task dependencies. Resource allocation includes application transformation, learning, and community detection. We use various fractal and statistical models to tame the impact of unknown security threats and incorporate that into the learning engines. The heterogeneous elements of S$^4$oC include reconfigurable cores, HW accelerators, and modules for security solutions such as cryptographic, non-binary numerical, and PUF that adapt to the existing environment. S$^4$oC manages the resources and reconfigures the elements in real-time, using security driven distributed RL. To optimize for security, the schedulers explore the best elements and reconfigurations onto which heterogeneous tasks should be mapped, according to application requirements and availability of S$^4$oC elements.